\documentclass[preprint,5p,twocolumn]{elsarticle}
\usepackage{lipsum}
\usepackage{tikz} \usetikzlibrary{positioning} \tikzset{font=\footnotesize}
\usepackage{standalone}
\usepackage{graphicx} 

\usepackage{color}
\usepackage{subcaption} 
\usepackage{makecell} 
\usepackage{amssymb, pifont}

\usepackage{amsmath,amssymb,amsfonts}
\usepackage[ruled, lined, linesnumbered, commentsnumbered, longend]{algorithm2e}
\usepackage{setspace}
\usepackage{amssymb}
\usepackage{tablefootnote}

\usepackage{graphicx}
\usepackage{textcomp}
\usepackage{xcolor}
\usepackage{soul}
\usepackage{enumerate}
\usepackage{comment}

\usepackage{multirow}
\usepackage{hyperref}
\usepackage{tabu}
\usepackage{float}
\usepackage{booktabs}
\usepackage{arydshln}
\usepackage[inline]{enumitem}

\usepackage{amsmath}
\usepackage{mathtools}
\usepackage{textcomp}
\usepackage{xurl}

\usepackage{threeparttable}

\definecolor{boristext}{rgb}{0.22, 0.44, 0.33}
\definecolor{boriscomments}{rgb}{0.83, 0.0, 0.0}
\definecolor{miguelcomments}{rgb}{0.5, 0, 0.8}
\definecolor{migueltext}{rgb}{0.42, 0.1, 0.9}

\begin{document}

\begin{frontmatter}

\title{Learning-Based Channel Access in Wi-Fi:\\A Multi-Armed Bandit Approach}

\journal{IEEE Transactions on Cognitive Communications and Networking}

\author[upf]{Miguel Casasnovas}
\author[upf]{Francesc Wilhelmi}
\author[supelec]{Richard Combes}
\author[agh]{Maksymilian Wojnar}
\author[agh]{Katarzyna Kosek-Szott}
\author[agh]{Szymon Szott}
\author[upf]{Anders Jonsson}
\author[uab]{Luis Esteve}
\author[upf]{Boris Bellalta}

\address[upf]{Universitat Pompeu Fabra, Barcelona, Spain}
\address[supelec]{CentraleSupélec, Paris, France.}
\address[agh]{AGH University of Krakow, Kraków, Poland.}
\address[uab]{Universitat Autònoma de Barcelona, Barcelona, Spain. \vspace{-2ex}}

\begin{abstract}
Due to its static protocol design, IEEE~802.11 (aka Wi-Fi) channel access lacks adaptability to address dynamic network conditions, resulting in inefficient spectrum utilization, unnecessary contention, and packet collisions. This paper investigates reinforcement learning~(RL) solutions to optimize Wi-Fi's medium access control~(MAC). In particular, a multi-armed bandit (MAB) framework is proposed for dynamic channel access (including both the primary channel and channel width) and contention window (CW) adjustment. In this setting, we study relevant learning design principles such as adopting joint or factorial action spaces (handled by a single agent (SA) and multiple agents (MA), respectively) and the importance of incorporating contextual information. Our simulation results show that cooperative MA architectures converge faster than their SA counterparts, as agents operate over smaller action spaces. Another key insight is that contextual MAB algorithms consistently outperform non-contextual ones, highlighting the value of leveraging side information in action selection. Moreover, in multi-player settings, results demonstrate that decentralized learners can achieve implicit coordination, although their greediness may degrade coexisting networks' performance and induce policy-chasing dynamics. Overall, these findings demonstrate that (contextual) MAB-based learning offers a practical and adaptive alternative to static IEEE~802.11 protocols, enabling more efficient and intelligent spectrum utilization.
\end{abstract}

\begin{keyword}
IEEE 802.11, Wi-Fi, channel allocation, machine learning, multi-armed bandits, multi-agent 
\end{keyword}

\end{frontmatter}



\section{Introduction}

Wi-Fi networks, standardized under the IEEE~802.11 family, operate in shared, unlicensed frequency bands using a distributed medium access mechanism known as the Distributed Coordination Function~(DCF)~\cite{geraci2025wi}. Due to the limited spectrum availability and the uncoordinated deployment of networks by users, multiple Basic Service Sets~(BSSs)\footnote{A BSS consists of a single access point~(AP) and one or more associated stations~(STAs).} often operate on overlapping channels and compete for the same frequency resources—commonly observed in residential buildings, office complexes, and shopping malls. In these heterogeneous and densely populated Overlapping BSS~(OBSS) environments, the current static channel access mechanisms of Wi-Fi lead to inefficient spectrum utilization, characterized by frequent collisions, excessive backoff periods, and significant throughput degradation. To address these challenges, it becomes essential to adapt channel access mechanisms to the diversity and dynamics of such scenarios. In particular, adaptive and learning-driven approaches hold great promise for enhancing spectrum efficiency and mitigating inter-BSS interference.

In this regard, reinforcement learning~(RL) has emerged as a suitable approach to optimize wireless networks~\cite{szott2022wi}, enabling devices to adapt through trial-and-error interactions with the environment by exploiting feedback from past actions. More specifically, multi-armed bandits~(MABs)---a stateless, model-free form of RL~\cite{bouneffouf2020survey}---are particularly suitable for Wi-Fi channel access because they focus on immediate transmission outcomes rather than long-term consequences,
avoid costly offline training phases that generalize poorly to
unseen conditions, and provide lightweight, sample-efficient online decision-making without requiring \textit{a priori} knowledge of the environment~\cite{slivkins2019introduction}.

This paper aims to contribute to the realization of future Machine Learning-Driven Radios~(MLDRs)~\cite{wilhelmi2024machine,szczech2025towards}. To this end, it proposes a novel MAB approach for Wi-Fi channel access, jointly learning the primary channel, channel width, and contention window~(CW) based on experienced performance rather than relying on DCF's static configurations. Under such a learning framework, we study fundamental design aspects, including the structure of the \textit{action space} (joint vs. factorial spaces, handled by a single-agent~(SA) and multiple agents~(MA), respectively) and the incorporation of \textit{contextual information} to improve learning efficiency. The main contributions of this paper are summarized as follows:  
\begin{enumerate}
    \item We propose a novel learning-driven Wi-Fi channel access mechanism, formulated as an online decision-making problem, in which learning-enabled APs adaptively select the primary channel, channel width, and CW based on observed network performance.
    \item We introduce SA and cooperative MA intra-AP architectures. In the SA architecture, learning APs include a single agent that optimizes all transmission parameters (channel and CW), providing complete control but requiring handling a combinatorial action space. In the MA architecture, learning APs include multiple agents, each focusing on a specific transmission parameter; thereby, problem tractability is improved by decomposing the action space into smaller subproblems. MA agents act independently yet align their learning objectives through a shared reward signal, enabling cooperative behavior.
    \item We evaluate both non-contextual (traditional) and contextual MAB formulations to assess the benefits of incorporating side information (e.g., channel occupancy, traffic load) in learning and adaptation. In the MA architecture, our Contextual MAB~(CMAB) formulation enables context sharing among agents to achieve \textit{implicit coordination} and overcome the lack of global optimization.
    \item We provide \texttt{WiPySim}\footnote{\url{https://github.com/miguelcUPF/WiPySim}},  a fully open-source, event-driven IEEE~802.11 simulator implemented in Python using SimPy. \texttt{WiPySim} supports both legacy and learning APs and includes flexible implementations of multiple MAB algorithms, enabling reproducible benchmarking of learning-based channel access mechanisms across configurable network scenarios.
    \item We evaluate the proposed learning approaches through comprehensive simulations in fully distributed single-player and multi-player settings, comparing their performance against legacy IEEE~802.11 baselines. Our results demonstrate the benefits of online learning-enabled methods (particularly contextual ones) over static legacy baselines under non-stationary wireless scenarios.
\end{enumerate}

The remainder of this paper is organized as follows: Section~\ref{sec:related_work} reviews related literature. Section~\ref{sec:architecture} formulates the proposed learning-driven channel access for Wi-Fi and details the different considered approaches. Section~\ref{sec:methodology} presents the system model and the evaluation methodology, including the simulation scenarios and hyperparameter tuning. Section~\ref{sec:results_1} provides a performance evaluation and Section~\ref{sec:discussion} discusses the results. Section~\ref{sec:conclusions} concludes the paper with key takeaways and future research directions.


\section{Related Work}\label{sec:related_work}

RL has recently been explored to redesign Wi-Fi MAC protocols beyond conventional parameter tuning. For instance, Pasandi et al.~\cite{pasandi2019challenges,pasandi2020mac} decomposed the MAC layer into modular building blocks (e.g., carrier sensing, backoff, contention window) and enabled a centralized agent to autonomously select the optimal combination, effectively automating protocol synthesis. Similarly, Keshtiarast et al.~\cite{keshtiarast2024ml,keshtiarast2024wireless} introduced a framework that dynamically selects, removes, or tunes such blocks under a centralized design in~\cite{keshtiarast2024ml} (where a single controller jointly configures all nodes) and a distributed design in~\cite{keshtiarast2024wireless} (where each node acts as an independent learner, locally optimizing its MAC stack). 

A substantial body of research has also focused on RL-based optimization of specific Wi-Fi functionalities. Learning-driven schemes have been proposed to dynamically adjust parameters such as the CW or backoff duration, reducing collisions and improving throughput compared to IEEE~802.11 DCF~\cite{chen2021study,kim2021performance,wydmanski2021contention,kumar2021adaptive,moon2021neuro}, or to allocate channels dynamically in dense deployments to mitigate contention and enhance efficiency~\cite{qi2020demand}. 

These studies typically rely on offline or episodic training paradigms (e.g., CTDE or DTE) and employ a single agent (either centralized or per-node) that jointly optimizes multiple MAC functionalities. In contrast, our approach performs fully online learning and further investigates multiple specialized agents per node, each dedicated to a specific building-block functionality. 

Complementary to general RL approaches, MABs have gained increasing attention for online wireless optimization owing to their simplicity, low computational cost, and adaptability. For channel access, Szczech et al.~\cite{szczech2025towards} leveraged MAB-based agents to tune MAC-layer parameters such as CW, RTS/CTS, and frame aggregation, reporting substantial gains over legacy IEEE~802.11 configurations. For channel bonding, Barrachina-Muñoz et al.~\cite{barrachina2021multi} applied MABs for primary-channel and channel-width selection, achieving notable improvements over static allocation strategies. For spatial reuse, Bardou et al.~\cite{bardou2021improving, bardou2023mitigating} proposed a centralized bandit that jointly optimizes transmit power and the OBSS/PD threshold (i.e., the sensitivity level used to decide whether transmissions from neighboring BSSs are treated as interference or ignored), significantly reducing starvation and improving fairness in dense deployments. Similarly, Wilhelmi et al.~\cite{wilhelmi2024coordinated} designed a coordinated bandit scheme leveraging IEEE~802.11 Multi-AP Coordination with joint reward structures (e.g., average or min-max) to improve spatial reuse across neighboring BSSs, while Wojnar et al.~\cite{wojnar2024ieee,wojnar2025coordinated} introduced hierarchical bandit models highlighting the efficiency and robustness of UCB-type algorithms. Moreover, Maghsudi et al.~\cite{maghsudi2014joint} and Wilhelmi et al.~\cite{wilhelmi2019collaborative} investigated multi-player bandit games for joint channel and power selection, demonstrating convergence to fair and stable equilibria even under distributed and selfish operation.

Recent efforts have further extended to contextual bandit formulations. For instance, Martínez et al.~\cite{martinez2023contextual} applied contextual bandits for network selection, incorporating traffic type as context and demonstrating that a polynomial extension of LinUCB outperforms non-contextual UCB under both stationary and non-stationary conditions. For spatial reuse, Iturria et al.~\cite{iturria2024cooperate} compared contextual and non-contextual bandits in cooperative and non-cooperative multi-player scenarios, showing that reward sharing among players greatly improves fairness and reduces starvation, and that cooperative contextual bandits are particularly effective in dynamic environments. 

Overall, while MABs have proven effective for optimizing Wi-Fi functionalities, their use for independently optimizing multiple MAC-layer components within the same device remains largely unexplored. This paper addresses this gap by investigating contextual and non-contextual bandit formulations for online, distributed channel-access optimization through a single-agent~(SA) approach, in which several MAC-layer building blocks are jointly optimized by a single learner, and a cooperative multi-agent~(MA) approach, in which multiple specialized agents concurrently optimize individual building blocks and, in the contextual case, coordinate implicitly through shared context. Our results provide new insights into the impact of action-space decomposition and contextual learning on Wi-Fi performance, establishing (C)MAB-based approaches as lightweight and adaptive alternatives to static IEEE~802.11 configuration mechanisms.

\section{Learning-Driven IEEE 802.11 Channel Access}\label{sec:architecture}

\begin{figure*}[t]
    \centering    \includegraphics[width=1\linewidth]{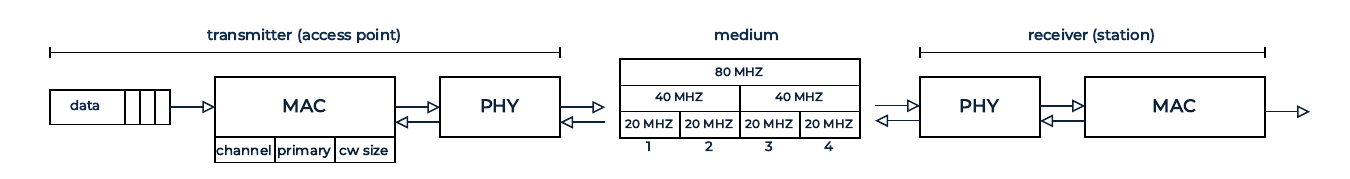}
    \caption{Transmitter--receiver 802.11 architecture.}
    \label{fig:mldr_arch}
\end{figure*}

\subsection{IEEE 802.11 Operation}

Wi-Fi devices communicate wirelessly following the IEEE~802.11 specification, which defines the Medium Access Control~(MAC) and Physical~(PHY) layers (see Fig.~\ref{fig:mldr_arch}). When an 802.11 transmitter (e.g., an AP) has data packets in its buffer, the MAC layer encapsulates them into data frames and performs channel access for their transmission to the intended destination (e.g., an associated STA). Once the wireless medium can be accessed, the PHY layer converts these frames into radio waves that propagate over specific frequency channels. At the receiver, the PHY layer captures these radio waves, the MAC reconstructs the original frames, and passes them to higher layers of the protocol stack.  

To improve spectral efficiency and data rates, recent IEEE~802.11 amendments support Channel Bonding~(CB), which aggregates contiguous 20~MHz basic channels into wider 40-, 80-, 160-, or 320-MHz bands. Within each bonded channel, one 20~MHz channel is designated as the \emph{primary (control) channel}, $p$, while the remaining ones serve as \emph{secondary channels}. 
In this work, we consider that basic channels can be aggregated according to either \emph{Static Channel Bonding (SCB)} or \emph{Dynamic Channel Bonding (DCB)}~\cite{barrachina2019dynamic}.

Wi-Fi relies on the Distributed Coordination Function~(DCF), a decentralized channel access protocol based on Carrier Sense Multiple Access with Collision Avoidance~(CSMA/CA) and the Binary Exponential Backoff~(BEB) mechanism, both designed to avoid or mitigate packet collisions in shared wireless environments. Through CSMA/CA, devices sense the medium to determine whether it is idle or busy (i.e., when the detected power from other transmissions exceeds a predefined threshold). Before initiating a transmission, the primary channel must remain idle for a DCF Inter-Frame Space~(DIFS) period. Once this condition is met, a randomized backoff timer---drawn uniformly from the range $[0, \text{CW}-1]$---is initiated and decrements with each idle time slot, ensuring distributed channel access. If the primary channel becomes busy before the counter reaches zero, the countdown pauses and resumes only after the channel is idle again for a complete DIFS period.  

Once the backoff counter reaches zero, the node may initiate transmission. Immediately prior to transmitting, all secondary channels are sensed for idleness over a Point Coordination Function Inter-Frame Space~(PIFS) period. Under SCB, the entire bonded channel is treated as a single unit: transmission is deferred if any channel in the bond is busy, meaning that all secondary channels must remain idle during the PIFS period. In contrast, DCB---introduced in IEEE~802.11ac (Wi-Fi~5)---adapts to instantaneous spectrum occupancy, enabling transmission on the primary channel and any contiguous idle secondary channels according to the IEEE~802.11 bonding rules. 
For instance, let $\mathcal{C} = \{c_1, c_2, \ldots, c_N\}$ denote the set of available 20-MHz basic channels, and let $C_{op} \subseteq \mathcal{C}$ represent the group of contiguous 20~MHz channels that the node configures for potential use (hereinafter referred to as the \emph{operational channel}). Under DCB, for $C_{op} = \{1p,2,3,4\}$ (with $p$ indicating the primary channel), a node may transmit on $\{1p,2\}$ even if $\{3\}$ and/or $\{4\}$ are busy, or only on $\{1p\}$ if $\{2\}$ is busy.

Finally, after each unsuccessful transmission, BEB doubles the CW, thereby reducing the probability of simultaneous retransmissions, albeit at the cost of longer waiting times.

\subsection{Problem Formulation}

\begin{figure*}[t]
    \centering
    \includegraphics[width=0.9\linewidth]{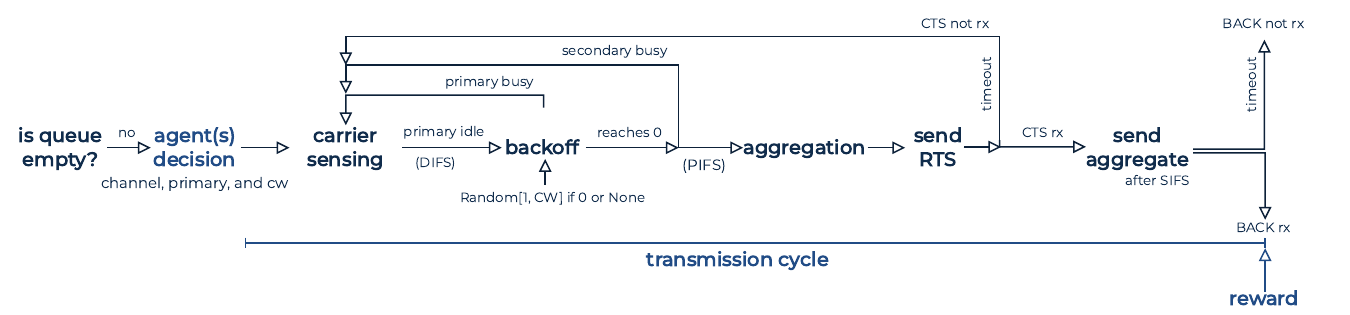}
    \caption{Learning access point transmission cycle.}
    \label{fig:tx_cycle}
\end{figure*}

We model Wi-Fi channel access as a sequential learning game in which, at the beginning of each round $t$, learning-enabled APs\footnote{Legacy (non-learning) APs may also be present, affecting the stochastic outcomes of the learning agents' actions.} select their transmission parameters, including the \emph{operational channel} $C_{op}$, \emph{primary channel} $p\in C_{op}$, and \emph{contention window (CW)}.  
To that end, we define learning rounds (asynchronous across APs) as \emph{transmission cycles}. As illustrated in Fig.~\ref{fig:tx_cycle}, each transmission cycle begins with carrier sensing and ends either upon reception of a Block ACK (or ACK) frame or after a timeout indicating transmission failure. To prevent persisting with ineffective configurations, a cycle may also be forcefully terminated if the AP fails to access the channel for a prolonged period (exceeding $D_{\max}$). Unlike standard IEEE~802.11, no BEB is applied, as
the learned CW is expected to be near-optimal.

To solve this sequential game, Wi-Fi channel access can be formulated as either a Multi-Armed Bandit (MAB) problem (aka $k$-armed bandit problem) or a Contextual Multi-Armed Bandit (CMAB) problem:

\begin{itemize}
    \item A MAB problem is defined by the tuple $\langle \mathcal{A}, \mathcal{R} \rangle$~\cite{rao_bandits}, where:
    \begin{itemize}
        \item $\mathcal{A}$ is the \emph{action space}, a finite set of $k$ discrete arms.
        \item $\mathcal{R} = \{\mathcal{R}^a\}_{a \in \mathcal{A}}$ is a family of unknown reward distributions, with $\mathcal{R}^a = \Pr[R_t \mid A_t = a]$.
    \end{itemize}
    \item A CMAB problem is defined by the tuple $\langle \mathcal{A}, \mathcal{X}, \mathcal{R} \rangle$~\cite{rao_bandits}, where:
    \begin{itemize}
        \item $\mathcal{A}$ is the \emph{action space}, a finite set of $k$ discrete arms.
        \item $\mathcal{X} \subseteq \mathbb{R}^d$ is the \emph{context space}, from which a $d$-dimensional feature vector $\mathbf{x}_t$ is observed at each round $t$.
        \item $\mathcal{R} = \{\mathcal{R}^a_\mathbf{x}\}_{a \in \mathcal{A}, \mathbf{x} \in \mathcal{X}}$ is a family of unknown context-conditional reward distributions, with $\mathcal{R}^a_\mathbf{x} = \Pr[R_t \mid A_t = a, \mathbf{X}_t =\mathbf{x}]$.
    \end{itemize}
\end{itemize}

The goal of a (C)MAB agent is to maximize the expected cumulative reward over $T$ rounds, or equivalently, to minimize the \emph{regret}, defined as the difference between the agent's cumulative reward and the reward it would have obtained by always selecting the optimal arm. To this end, the agent sequentially selects an action (or arm) $a_t \in \mathcal{A}$ at each round $t \in \{1, \dots, T\}$. In CMABs, this selection is conditioned on the observed context $\mathbf{x}_t \in \mathcal{X}$. At the end of each round, the environment reveals a reward $r_t$, which may be stochastic or non-stochastic depending on the problem. This reward is then used by the agent to refine its action-selection policy $\pi$, guiding future decisions.

\subsection{Action Design} 

At each round, the learning AP selects an action vector $a_t = (a^{\text{ch}}, a^{\text{p}}, a^{\text{cw}})$, where:
\begin{itemize}
    \item $a^{\text{ch}} \in \{ \{1\}, \{2\}, \{3\}, \{4\}, \{1,2\}, \{3,4\}, \{1,2,3,4\} \}$ is the set of 20~MHz channels that the AP may use for transmission ($C_{op}$), and thus determines the channel width (20/40/80~MHz).\footnote{Throughout this paper, we assume that the set of available basic channels is $\mathcal{C} = \{1,2,3,4\}$ and that receivers continuously monitor all basic channels when idle, allowing transmitters to switch to any channel group on a per-transmission basis.}
    \item $a^{\text{p}} \in a^{\text{ch}}$ is the selected primary channel ($p$), known by the clients prior to transmission.
    \item $a^{\text{cw}} \in \{ 2^{i+4} \mid i = 0, \dots, 6 \}$ is the selected CW size, covering the standard IEEE~802.11 range~\cite{szczech2025towards}.
\end{itemize}

Depending on the learning architecture, actions are selected either jointly or decomposed across agents:
\begin{itemize}
    \item Single-Agent (SA): A single agent jointly selects the full transmission configuration, yielding the joint action vector $a$. Accounting for the conditional dependence of the primary channel on the selected operational channel, the SA architecture has a joint action space of size $|\mathcal{A}^{\text{SA}}| = 84$.
    \item Multi-Agent (MA): Three agents operate independently and sequentially, each responsible for one parameter: channel $\Rightarrow$ primary $\Rightarrow$ CW. This decomposition improves tractability, as each agent is responsible for its respective action set, with sizes $|\mathcal{A}^{\text{ch}}| = 7$, $|\mathcal{A}^{\text{p}}| \leq 4$ (depending on the selected operational channel), and $|\mathcal{A}^{\text{cw}}| = 7$.
\end{itemize}

\subsection{Reward Design} 

We consider a scalar reward $r\in[0, 1]$, computed from the transmission cycle duration, $D$, using min--max normalization and clipping:
\begin{equation*}
r = 1-\text{clip}_{[0,1]}\!\left(\frac{D - D_{\min}}{D_{\max} - D_{\min}} \right) = \text{clip}_{[0,1]}\!\left(\frac{D_{\max} - D }{D_{\max} - D_{\min}} \right)\!,
\end{equation*}
with $D_{\min} = 0$~ms and $D_{\max} = 10$~ms (approx. 80th percentile latency of a Wi-Fi hop~\cite{sui2016characterizing}). This design encourages faster and more efficient medium access. In the MA architecture, all agents share the same scalar reward to promote cooperation.


\subsection{Context Design (for contextual bandits only)} 
In our design, each CMAB-based agent incorporates side information from both the device and its surrounding wireless environment to support informed decision-making. In particular, each agent observes a normalized context vector $\mathbf{x} \in [0,1]^d$ tailored to its decision domain (see Table.~\ref{tab:context_mapping}), including the following features: 
\begin{itemize}
    \item \textit{F1. Channel occupancy ratios} ($[0,1]^4$): fraction of time each 20~MHz basic channel is sensed as busy (excluding transmissions from the AP itself and its associated STAs), averaged over a 100~ms sliding window. 
    \item \textit{F2. Binary channel busy/idle flags} ($\{0,1\}^4$): instantaneous activity state for each 20~MHz channel, where 0 indicates idle and 1 indicates busy.
    \item \textit{F3. Transmission queue utilization} ($[0,1]$): fraction of the MAC-layer transmission queue currently in use. 
    \item \textit{F4. Selected operational channel} ($\{0,1\}^4$): one-hot encoded representation of the selected operational channel. 
    \item \textit{F5. Selected primary channel} ($\{0,1\}^4$): one-hot encoded representation of the selected primary channel.  
\end{itemize}

This design enables \textit{implicit coordination} in the MA architecture, as some features are shared and later agents incorporate prior agents' decisions into their context.


\subsection{Exploration-Exploitation Strategies} \label{sec:strategies}

To solve a MAB problem, an agent must balance \emph{exploration} (selecting less-known actions to improve reward estimates) and \emph{exploitation} (choosing actions with high expected reward). 
This paper evaluates two different state-of-the-art action-selection algorithms: Upper Confidence Bound~(UCB)~\cite{auer2002finite} and linear UCB (LinUCB)~\cite{li2010contextual}.

\textit{UCB} follows the principle of optimism in the face of uncertainty, relying on confidence bounds that tighten as actions are sampled more frequently. In particular, at each round $t$, UCB($\alpha$) selects an action $a\in \mathcal{A}$ according to:
\begin{equation*}
    a_t = \arg\max_{a \in \mathcal{A}} \left[ \hat{\mu}_a(t) + \sqrt{\frac{\alpha \ln t}{2 N_a(t)}} \right],
\end{equation*}
where $\hat{\mu}_a(t)$ is the empirical mean reward of action $a$, $N_a(t)$ its selection count, and $\alpha > 0$ is a tunable parameter that controls exploration (e.g., UCB1 uses $\alpha=4$). UCB does not use contextual information and assumes stationary reward distributions. 

\textit{LinUCB} (disjoint) is a contextual adaptation of the classical UCB algorithm. It assumes a disjoint linear model\footnote{This model is called \textit{disjoint} because the parameters are not shared among arms; each action a has its own parameter vector $\boldsymbol{\theta}_a^*$.} of expected rewards $\mathbb{E}[r_t \mid \mathbf{x}_t] = \mathbf{x}_t^\top \boldsymbol{\theta}_a^*$ and selects actions according to:
\begin{equation*}
    a_t = \arg\max_{a \in \mathcal{A}} \left[ \hat{\boldsymbol{\theta}}_a^\top \mathbf{x}_t + \alpha \sqrt{\mathbf{x}_t^\top \mathbf{A}_a^{-1} \mathbf{x}_t} \right],
\end{equation*}
where $\hat{\boldsymbol{\theta}}_a$ is a ridge regression estimate of the true parameter vector $\boldsymbol{\theta}_a^*$, and $\mathbf{A}_a = \mathbf{D}_a^\top \mathbf{D}_a + \mathbf{I}_d$, with $\mathbf{D}_a \in \mathbb{R}^{t \times d}$ denoting the design matrix---whose $i$-th row is the context vector observed when action $a$ was selected at time $i$---and $\mathbf{I}_d$ being the $d \times d$ identity matrix.
It can be extended to handle non-stationarity; for instance, Sliding-Window LinUCB~(SW-LinUCB)~\cite{gutowski2019global} incorporates a discount factor that penalizes frequently selected actions over a sliding window of $w$ rounds.\footnote{Initially, we considered SW-LinUCB; nevertheless, hyperparameter tuning revealed that the sliding window consistently degraded performance, as it penalized frequently selected high-reward actions and increased exploration of suboptimal ones. 
}


\begin{table}[t]
\centering
\caption{Contextual features used by the single-agent (SA) and multi-agent (MA) CMAB architectures.}
\footnotesize
\label{tab:context_mapping}
\begin{tabular}{@{}l|c|ccc@{}}
                 &            & \textbf{grp} & \textbf{pri} & \textbf{cw} \\
\toprule
\textbf{Feature} & \textbf{SA} & \multicolumn{3}{c}{\textbf{MA}} \\
\midrule
F1. Occupancy ratio     & \checkmark & \checkmark & \checkmark & \checkmark \\
F2. Busy/idle flags     & \checkmark & \checkmark & \checkmark & \checkmark \\
F3. Queue utilization   & \checkmark & \checkmark & $\times$ & \checkmark \\
F4. Operational channel         & $\times$ & $\times$     & \checkmark & \checkmark \\
F5. Primary channel    & $\times$ & $\times$     & $\times$     & \checkmark \\
\bottomrule
\multicolumn{1}{r}{$d=$} & \multicolumn{1}{c}{9} & \multicolumn{1}{c}{9} & \multicolumn{1}{c}{12} & \multicolumn{1}{c}{17}
\end{tabular}
\end{table}


\section{Evaluation Methodology}
\label{sec:methodology}

The (C)MAB solutions described in Section~\ref{sec:architecture} are evaluated under both the SA and MA approaches through simulations involving multiple BSSs competing for medium access in a \emph{fully decentralized and non-cooperative} manner. For MA, all agents are homogeneous, operating independently but using the same algorithms and hyperparameters. In addition to learning-enabled approaches, static, legacy IEEE~802.11 performance is provided as a baseline.

\subsection{Scenarios}

\begin{figure*}[t!]
\centering
    \begin{subfigure}[t]{0.39\linewidth}
        \centering
        \includegraphics[width=0.9\linewidth]{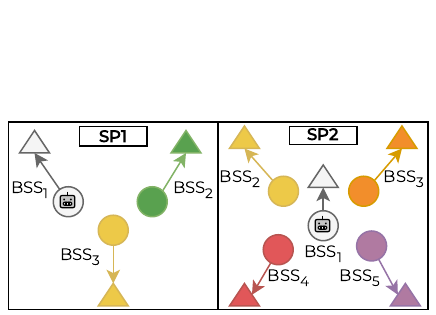}
        \caption{Single-player (SP) scenarios.}
        \label{fig:sp_scenarios}
    \end{subfigure}
    \begin{subfigure}[t]{0.58\linewidth}
        \centering
            \includegraphics[width=0.9\linewidth]{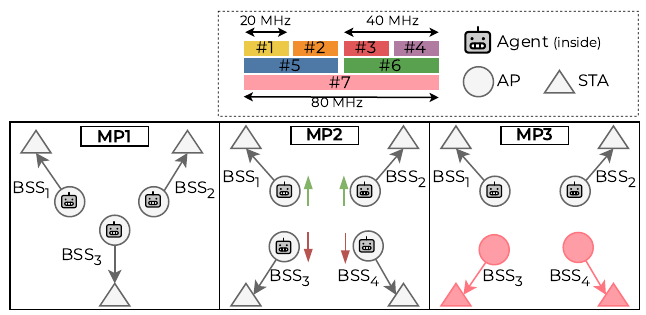}
        \caption{Multi-player (MP) scenarios.}
        \label{fig:mp_scenarios}
    \end{subfigure}
    \caption{Schematic representation of the considered evaluation scenarios.}
    \label{fig:scenarios}
\end{figure*}

The proposed mechanisms are assessed in $(i)$ a \emph{single-player~(SP) game}, where only one AP employs learning, and $(ii)$ a \emph{selfish multi-player~(MP) game}, where multiple self-interested learning APs coexist. The SP and MP scenarios, both under an SCB-constrained system, are analyzed in Sections~\ref{sec:results_1} and~\ref{sec:results_2}, respectively. The DCB case is examined in Section~\ref{sec:results_3}. In all cases, legacy BSSs rely on static channel assignments and standard IEEE~802.11 access mechanisms, with their primary channel set to the lowest-indexed 20~MHz channel within the assigned channel group.

SP scenarios, illustrated in Fig.~\ref{fig:sp_scenarios}, evaluate the effectiveness and adaptability of the considered algorithm--architecture pairs:
\begin{itemize}
  \item \emph{SP1 -- One empty channel:} One learning-enabled BSS (BSS$_1$) coexists with two legacy BSSs (BSS$_2$ and BSS$_3$). BSS$_2$ is in channels~\#6:$\{3,4\}$ (40~MHz) and BSS$_3$ in channel~\#1:$\{1\}$ (20~MHz). Channel \#2:$\{2\}$ (20~MHz) remains unassigned, offering exclusive access to the learning AP. All APs have full-buffer traffic. 
  \item \emph{SP2 -- Dynamic OBSS traffic:} One learning-enabled BSS (BSS$_1$) coexists with four other legacy BSSs (BSS$_2$ to BSS$_5$), each operating on a distinct non-overlapping 20~MHz channel. The learning AP has full-buffer traffic, while legacy APs generate dynamic, non-saturated traffic following Poisson (packets generated individually with exponentially distributed inter-packet times), Bursty (packets generated in bursts with exponentially distributed inter-burst times), or Virtual Reality~(VR, packets generated in bursts according to a fixed frame rate) models, assigned randomly. The scenario is structured into four sequential \emph{intervals}, each lasting 15 seconds (60 seconds in total). At the beginning of each interval, the load of each legacy AP is updated, generally remaining between 80--90\% of its maximum theoretical goodput. To evaluate adaptability, in each of the first three intervals, a distinct legacy AP is randomly selected (without repetition) to temporarily operate at reduced load~(10--20\% of its maximum theoretical goodput). In the fourth (and final) interval, one of the three previously selected APs operates again under reduced load. Thus, in each interval, a temporarily under-loaded channel provides a transient opportunity for increased medium access to the learning AP. 
\end{itemize}

MP scenarios, illustrated in Fig.~\ref{fig:mp_scenarios}, introduce additional complexity from adversarial inter-agent interactions and induced \emph{non-stationarity} due to concurrent policy updates: 
\begin{itemize}
  \item \emph{MP1 -- Learning under full-buffer conditions:} Three learning-enabled BSSs (BSS$_1$ to BSS$_3$) operate under full-buffer traffic.  
  \item \emph{MP2 -- Learning under heterogeneous load distributions:} Four learning-enabled BSSs coexist. BSS$_1$ and BSS$_2$ operate under high traffic (full-buffer) and BSS$_3$ and BSS$_4$, under low traffic (20--40\% of maximum theoretical goodput). 
  \item \emph{MP3 -- Learning vs. Legacy coexistence:} Two learning-enabled BSSs (BSS$_1$ and BSS$_2$) coexist with two legacy BSSs (BSS$_3$ and BSS$_4$) that operate in 80~MHz channels. All BSSs generate heterogeneous, non-full-buffer traffic with offered loads ranging from 60\% to 90\% of the maximum theoretical goodput.  
\end{itemize}

\subsection{Simulation Setup and Metrics} 

Simulations are conducted using \texttt{WiPySim}, a custom, open-source, event-driven IEEE~802.11 simulator developed in Python. The main simulation parameters are summarized in Table~\ref{tab:simparams}.
Each simulation runs for 60~seconds and is repeated across twenty independent trials. Device positions remain fixed across trials, with all APs and STAs located within a $10 \times 10 \times 2$~m area, ensuring MCS~11 is achievable and all devices remain within mutual coverage. Each BSS comprises a single AP--STA pair, and only downlink traffic is considered. For all simulations, the set of available 20~MHz basic channels is $\mathcal{C} = \{1,2,3,4\}$, and thus the set of valid operational channels (i.e., all subsets $C_{op} \subseteq \mathcal{C}$ that a BSS may use) is
$\mathcal{O} =\{\{1\}, \{2\}, \{3\}, \{4\}, \{1,2\}, \{3,4\}, \{1,2,3,4\}\}$. Each $C_{op} \in \mathcal{O}$ corresponds to a possible channel configuration and is labeled as \#1--\#7, in the same order as listed in $\mathcal{O}$.

Results are reported for each learning-enabled BSS, either aggregated across trials or for a single representative trial, selected as the one most clearly illustrating the (consistently observed) phenomenon discussed.
Performance is measured over the entire simulation duration, except in Scenario~SP2, where results are also reported separately across intervals to capture temporal dynamics. A 2~second burn-in period is excluded from most of the evaluations---both global and interval-based---to avoid bias from transients during the initial learning phase. Nonetheless, temporal evolution plots include these transient phases to report the relative learning speed and stability.

All collected data, including simulation statistics and performance metrics, is stored in a publicly available dataset on Zenodo.\footnote{\url{https://doi.org/10.5281/zenodo.17347998}}  
Among the collected metrics, the following are used in this paper to quantify the performance of each BSS:
\begin{itemize}
  \item Goodput ($\Gamma_i$): rate of successfully delivered application-layer data in BSS$_i$, capturing both effective throughput and overall transmission efficiency. Instantaneous goodput is sampled at the reception of each data frame, and the average is computed as a time-weighted mean to account for non-uniform sampling intervals.
  \item Delay ($d_i$): per-packet delay in BSS$_i$, measured from packet generation to reception, including queuing, backoff, and (re)transmission times (up to 7 retransmissions).
\end{itemize}
Additionally, to evaluate the distribution of resources among BSSs, this paper considers goodput fairness, which is computed offline:
\begin{itemize}
\item Fairness ($\mathcal{J}$): goodput fairness across a set of BSSs, measured using Jain’s fairness index~\cite{jain1984quantitative}. For a set of BSSs, $\mathcal{S}$, the fairness index is defined as
\begin{equation*}
    \mathcal{J}_{\mathcal{S}} =
    \frac{\left(\sum_{i \in \mathcal{S}} \overline{\Gamma}_i \right)^2}
    {|\mathcal{S}| \sum_{i \in \mathcal{S}} \overline{\Gamma}_i^2},
\end{equation*}
where $\overline{\Gamma}_i$ denotes the average goodput achieved by BSS$_i$. The index ranges from~0 (unequal) to~1 (perfect fairness).
\end{itemize}
\subsection{Hyperparameter Tuning}

Algorithm hyperparameters are optimized independently for each algorithm--architecture pair using \texttt{Optuna}\footnote{\url{https://optuna.org/}}, following a scenario-based methodology similar to~\cite{wojnar2024ieee}. Each optimization evaluates 100 candidate configurations across diverse, reproducible deployments that combine different numbers of coexisting BSSs (2, 3, or 4) and simulation durations~(1, 2, 4, or 8~seconds).
In each deployment, a single learning-enabled AP coexists with multiple legacy APs operating under SCB, capturing representative interactions in multi-AP environments. Each legacy AP selects a operational channel $C_{op}\in \mathcal{O}$ uniformly at random and sets its primary channel to the lowest-indexed 20~MHz channel in $C_{op}$. APs and STAs are randomly placed in a 10$\times$10$\times$2~m area, and traffic models (Poisson, Bursty, VR, or full-buffer) and parameters (e.g., load, inter-arrival times) are assigned uniformly at random per AP. 
The hyperparameter search ranges are $\alpha \in (1.0,,10.0)$ for UCB and $\alpha \in (0.2,,20.0)$ for LinUCB, and the configuration yielding the highest average reward across deployments is selected: $\alpha = 1.09$ (SA) and $1.14$ (MA) for UCB, and $\alpha = 0.52$ (SA) and $0.50$ (MA) for LinUCB.

\begin{table}[t]
\centering
\caption{Simulation parameter settings.}
\setlength{\tabcolsep}{4pt}
\footnotesize
\label{tab:simparams}
\begin{tabular}{@{}llll@{}}
\toprule
\textbf{Parameter} & \textbf{Value} & \textbf{Parameter} & \textbf{Value} \\
\midrule
Carrier frequency & 5 GHz & MAC/PHY & IEEE 802.11ax\\
No. basic channels & 4  & Bandwidth & 20/40/80 MHz \\
MCS index & 0--11 & Spatial streams & 2 \\
Tx power & 20 dBm & Gain (Tx/Rx) & 0 dB / 0 dB \\
RTS/CTS & Enabled & Tx queue size & 500 packets \\
max A-MPDU size & 65,535 B & CW (min/max) & 16 / 1024 \\
Retry limit & 7 & Channel bonding & Static/Dynamic \\
Path loss\tablefootnote{The free-space log-distance path loss model is used, assuming a reference distance of 1~m.} exponent & 4 & Packet error rate & 0.1 \\
\bottomrule
\end{tabular}
\end{table}

\section{Results: Single-Player Game}\label{sec:results_1}

This section presents the simulation results for the SP game under SCB. In this setting, the AP from BSS$_1$ is the only learning-enabled node and coexists with several legacy APs. For benchmarking, IEEE~802.11 legacy baselines with all possible static channel assignments for BSS$_1$ are also included.

\subsection{One empty channel (Scenario SP1)}

\begin{figure}[t!]
\centering
    \begin{subfigure}[t]{\linewidth}
        \centering
        \includegraphics[width=\linewidth]{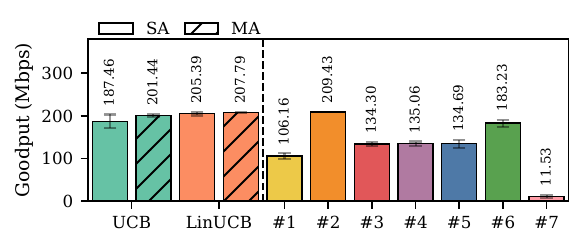}
        \caption{Goodput.}
        \label{fig:res_a_goodput}
    \end{subfigure}%
    \hfill
    \begin{subfigure}[t]{\linewidth}
        \centering
            \includegraphics[width=\linewidth]{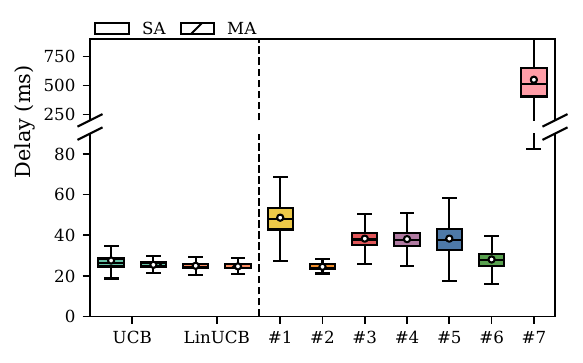}
        \caption{Packet delay.}
        \label{fig:res_a_del}
    \end{subfigure}
    \caption{Scenario~SP1 results for BSS$_1$ under SCB across all trials, comparing each algorithm--architecture pair and static non-learning baseline. Error bars show the mean and standard deviation across trials, while boxplots represent the distribution of all values collected across trials.}
    \label{fig:res_a}
\end{figure}

\begin{figure}[t!]
\centering
    \begin{subfigure}[t]{\linewidth}
        \centering
        \includegraphics[width=\linewidth]{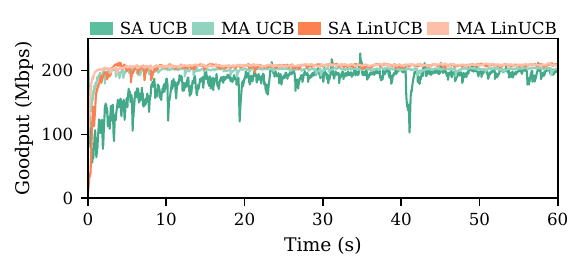}
        \caption{Scenario~SP1.}
        \label{fig:a_evo}
    \end{subfigure}%
    \hfill
    \begin{subfigure}[t]{\linewidth}
        \centering
        \includegraphics[width=\linewidth]{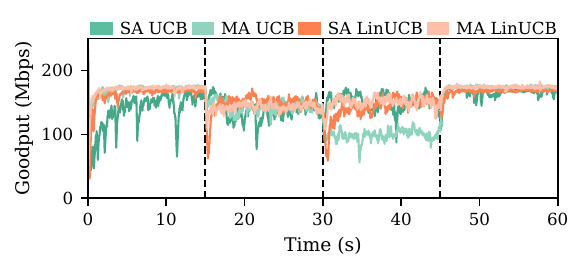}
        \caption{Scenario~SP2.}
        \label{fig:b_evo}
    \end{subfigure}

    \caption{Temporal evolution of goodput in representative trials under SCB for each algorithm--architecture pair.}
    \label{fig:goodput_evo_both}
\end{figure}

Figures~\ref{fig:res_a_goodput} and~\ref{fig:res_a_del} report the goodput (in Mbps) and delay~(in ms) experienced by BSS$_1$ under the considered approaches. Learning-based solutions are shown on the left-hand side of the plots, while legacy IEEE~802.11 performance is displayed on the right. In addition, Fig.~\ref{fig:a_evo} illustrates the temporal evolution of goodput for the learning approaches in a representative trial.

\noindent\textit{\textbf{IEEE 802.11 legacy operation:}}
The high performance differences across channel configurations, illustrated in Figures~\ref{fig:res_a_goodput} and~\ref{fig:res_a_del} (right), demonstrate that channel allocation critically determines performance. For BSS$_1$, the contention-free channel \#2:$\{2\}$ provides the highest performance across trials ($\overline{\Gamma}_1 =209.4$~Mbps, $\overline{d}_1 =24.4$~ms). In contrast, using channel \#6:$\{3,4\}$ (fully overlapping BSS$_2$), provides the second-best performance ($\overline{\Gamma}_1 =183.2$~Mbps, $\overline{d}_1 =28.0$~ms), whereas using \#5:$\{1,2\}$ (partially overlapping BSS$_3$ on channel \#1:$\{1\}$) yields substantially lower performance ($\overline{\Gamma}_1 =134.7$~Mbps, $\overline{d}_1 =38.3$~ms), indicating that partial overlaps can be more detrimental than full overlaps under SCB. 
Allocations \#3:$\{3\}$ ($\overline{\Gamma}_1 =134.3$~Mbps, $\overline{d}_1 =38.3$~ms) and \#4:$\{4\}$ ($\overline{\Gamma}_1 =135.1$~Mbps, $\overline{d}_1 =38.0$~ms) achieve similar performance to \#5:$\{1,2\}$ despite their smaller bandwidth and, consequently, lower nominal transmission rates. This occurs because channels \#3:$\{3\}$ and \#4:$\{4\}$ overlap with a 40~MHz full-buffer contender, whereas \#5:$\{1,2\}$ overlaps with a 20~MHz full-buffer contender. The 40~MHz contender completes its transmissions faster and thus releases the medium more frequently, providing the 20~MHz allocations more frequent access opportunities and effectively compensating for their lower transmission rates. Indeed, Fig.~\ref{fig:res_a_del} shows that \#5:$\{1,2\}$ achieves a smaller 25th percentile delay due to higher transmission rates, but exhibits a larger 75th percentile delay due to increased contention with a slower contender.  
Other channel allocations substantially degrade performance, e.g., \#1:$\{1\}$ ($\overline{\Gamma}_1 = 106.1$~Mbps, $\overline{d}_1 = 48.6$~ms) and \#7:$\{1,2,3,4\}$ ($\overline{\Gamma}_1 = 11.5$~Mbps, $\overline{d}_1 = 550.5$~ms), underscoring the detrimental impact of misconfiguration. 
Importantly, while not depicted in the figures, channel \#2:$\{2\}$ also maximizes goodput for legacy BSSs ($\overline{\Gamma}_2 = 360.6$~Mbps; $\overline{\Gamma}_3 = 209.4$~Mbps). In contrast, poor allocations can substantially reduce their performance (e.g., \#1:$\{1\}$ reduces BSS$_3$ to $105.8$~Mbps, and \#6:$\{3,4\}$ reduces BSS$_2$ to $183.8$~Mbps).  

\noindent\textit{\textbf{Learning algorithms:}} 
As illustrated in Figures~\ref{fig:res_a_goodput} and~\ref{fig:res_a_del} (left),
 UCB approaches the performance of the best baseline (\#2:$\{2\}$) while clearly outperforming the second-best baseline (\#6:$\{3,4\}$), particularly in the MA case (SA: $\overline{\Gamma}_1 = 187.5$~Mbps, $\overline{d}_1 = 27.5$~ms; MA: $\overline{\Gamma}_1 = 201.4$~Mbps, $\overline{d}_1 = 25.4$~ms). 
 Indeed, MA~UCB almost exclusively selects the contention-free allocation \#2:$\{2\}$ ($99.7\%$ of selections across trials), while SA~UCB does so less frequently ($76.5\%$), with $18.3\%$ of selections on \#6:$\{3,4\}$, explaining its comparatively lower performance.    
MA~UCB not only outperforms SA~UCB but also converges faster, as shown in Fig.~\ref{fig:a_evo}, a behavior consistently observed across trials. 
However, despite consistently converging to the optimal channel, MA~UCB exhibits inherent coordination limitations that lead to suboptimal CW values (e.g., CW~32 in $99.7\%$ of selections). 
This limitation stems from the combination of the standard UCB initialization procedure and the shared reward structure used in the MA setting. Each action is initially pulled once before UCB-based selection begins, proceeding sequentially across the action space. Consequently, in the MA case, actions with the same index across agents are sampled simultaneously (e.g., the second channel option with the second CW option). Because all agents share a common, non-separable scalar reward, their empirical estimates evolve identically, causing their actions to become tightly coupled over time. As a result, agents face a \emph{credit assignment problem} (i.e., the inability to determine which agent’s action contributed to the observed outcome), which hinders independent exploration of the joint action space.  
Notably, randomized initialization would only yield different fixed action associations, without addressing the underlying coupling problem inherent to vanilla UCB under shared rewards. 

On the other hand, as depicted in Figures~\ref{fig:res_a_goodput} and~\ref{fig:res_a_del} (left), 
LinUCB attains the highest performance across trials, closely matching the contention-free baseline (SA: $\overline{\Gamma}_1 = 205.4$~Mbps, $\overline{d}_1 = 24.9$~ms; MA: $\overline{\Gamma}_1 = 207.5$~Mbps, $\overline{d}_1 = 24.6$~ms). Both SA and MA variants almost exclusively select \#2:$\{2\}$ (SA: $98.8\%$; MA: $100\%$) and consistently use small CWs (SA: CW~16, $69.1\%$ and CW~32, $26.2\%$; MA: CW~16, $86.1\%$ and CW~32, $11.2\%$), maximizing access opportunities. The MA variant again outperforms its SA counterpart and, as illustrated in Fig.~\ref{fig:a_evo}, it converges significantly faster, a recurring behavior across trials. MA~LinUCB avoids the joint-action coupling observed in MA~UCB because each agent conditions its action selection on independent contextual information, providing distinct estimates, even under a shared reward, and preventing the credit assignment problem.

\noindent\textit{\textbf{Discussion:}}
Scenario~SP1 confirms that the considered learning mechanisms can successfully identify and exploit the contention-free channel. Both UCB and LinUCB achieve performance comparable to the optimal baseline; however, LinUCB attains notably higher performance and faster convergence, demonstrating the benefits of incorporating contextual information into the decision process. Moreover, MA~UCB exhibits coupling limitations arising from the shared-reward setting, whereas MA~LinUCB avoids them by conditioning each agent’s action selection on independent contextual information. Overall, SA and MA architectures achieve comparable steady-state performance, though MA consistently converges faster across trials. Finally, since learned configurations favor contention-free operation, learning mechanisms did not negatively impact neighboring legacy BSSs, preserving their performance (e.g., MA~LinUCB: $\overline{\Gamma}_2=360.5$~Mbps; $\overline{\Gamma}_3 = 209.4$~Mbps). 

\vspace{0.2cm}
\noindent\fbox{%
    \parbox{0.97\columnwidth}{%
    From this point, given the strong correlation observed between goodput and packet delay, subsequent results focus exclusively on goodput. 
    }%
}

\subsection{Dynamic OBSS traffic (Scenario~SP2)} 

\begin{figure}[t!]
\centering
    \begin{subfigure}[t]{\linewidth}
        \centering
        \includegraphics[width=\linewidth]{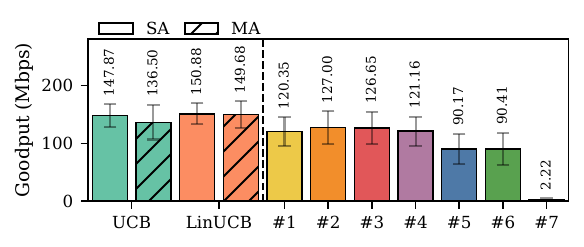}
        \caption{Aggregate across trials.}
        \label{fig:res_b_goodput}
    \end{subfigure}%
    \hfill
    \begin{subfigure}[t]{\linewidth}
        \centering
        \includegraphics[width=\linewidth]{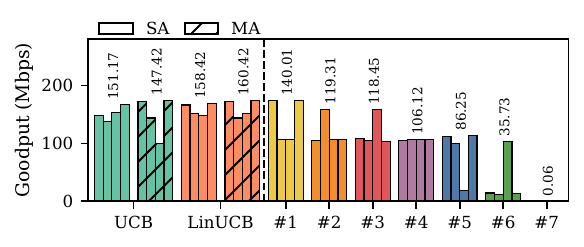}
        \caption{Representative trial (sub-bars correspond to 15-second intervals).}
        \label{fig:res_b_goodput_intervals}
    \end{subfigure}
    \caption{Scenario~SP2 goodput results for BSS$_1$ under SCB, comparing each algorithm--architecture pair and static non-learning baseline.}
    \label{fig:res_b_combined}
\end{figure}

Fig.~\ref{fig:res_b_goodput} reports the goodput achieved by the different learning methods and IEEE~802.11 baselines across all trials. Fig.~\ref{fig:res_b_goodput_intervals}, in contrast, presents the goodput observed in each 15-second interval of the simulation for a representative trial. Finally, Fig.~\ref{fig:b_evo} illustrates the temporal evolution of goodput for the learning approaches in the same representative trial.

\noindent \textit{\textbf{IEEE 802.11 legacy operation:}} 
As shown in Fig.~\ref{fig:res_b_goodput} (right), 
under SCB, the 80~MHz allocation \#7:$\{1,2,3,4\}$ yields the weakest performance across trials ($\overline{\Gamma}_1 =2.2$~Mbps), as increased contention dominates the potential gains of higher nominal transmission rates. Static 20~MHz allocations ($\#1:\{1\}$, \#2:$\{2\}$, \#3:$\{3\}$, and \#4:$\{4\}$) achieve relatively uniform performance across trials, since in each interval one legacy BSS reduces its load and temporarily decreases contention on the corresponding channel. These allocations are thus occasionally optimal, but their lack of adaptability prevents systematic exploitation of favorable conditions, as shown in Fig.~\ref{fig:res_b_goodput_intervals} (right). Indeed, the best-performing 20~MHz allocation varies across intervals, as observed for the representative trial: channel~\#1:$\{1\}$ in the first ($\overline{\Gamma}_1 = 173.5$~Mbps), \#2:$\{2\}$ in the second ($\overline{\Gamma}_1 = 159.1$~Mbps), \#3:$\{3\}$ in the third ($\overline{\Gamma}_1 = 158.0$~Mbps), and \#1:$\{1\}$ again in the fourth ($\overline{\Gamma}_1 = 173.7$~Mbps). In contrast, 40~MHz allocations exhibit significant degradation whenever none of their composing channels carries a lower load (e.g., for the representative trial, \#5:$\{1,2\}$ in the third interval: $\overline{\Gamma}_1 = 19.1$~Mbps). The 80~MHz allocation consistently performs worst across intervals (as low as $\overline{\Gamma}_1 = 0.1$~Mbps), due to minimal channel access opportunities under SCB. These results confirm that effective operation in dynamic conditions requires adaptive strategies capable of identifying and exploiting the temporarily underloaded channel, as the optimal allocation changes over time.  

\noindent\textit{\textbf{Learning algorithms:}} 
As illustrated in Fig.~\ref{fig:res_b_goodput} (left),
 UCB achieves high performance in both architectures, clearly surpassing the best-performing baselines (SA: $\overline{\Gamma}_1 = 147.9$~Mbps; MA: $\overline{\Gamma}_1 = 136.5$~Mbps). 
In particular, as depicted in Fig.~\ref{fig:res_b_goodput_intervals} (left),
SA~UCB sustains high performance across intervals. Indeed, it consistently converges to the interval-wise optimal channels, as observed for the representative trial: channel~\#1:$\{1\}$ is selected in $86.4\%$ of decisions during the first interval, \#2:$\{2\}$ in $86.5\%$ during the second, \#3:$\{3\}$ in $96.7\%$ during the third, and \#1:$\{1\}$ again in $98.5\%$ during the fourth. Conversely, in the same representative trial, MA~UCB, demonstrates faster convergence in the first interval but underperforms in the third (see Fig.~\ref{fig:b_evo}). Indeed, although it also converges to optimal allocations in most intervals (\#1:$\{1\}$: $99.4\%$ in the first, \#2:$\{2\}$: $79.3\%$ in the second, \#1:$\{1\}$: $100\%$ in the fourth), in the third it remains on channel~\#1:$\{1\}$ instead of switching to \#3:$\{3\}$. This reflects the coordination issue discussed in Scenario~SP1: agents select specific joint action combinations (e.g., channel~\#1:$\{1\}$ with CW~16, channel~\#3:$\{3\}$ with CW~64). As a result, in the third interval, channel~\#3:$\{3\}$ is not selected because the learned joint action yields lower observed rewards due to its associated higher CW. This behavior is consistently observed across all trials whenever channels~\#3:$\{3\}$ or \#4:$\{4\}$ constitute the interval-wise optimum, due to their high associated CWs.

On the other hand, as depicted in Fig.~\ref{fig:res_b_goodput} (left),
LinUCB achieves the best performance across trials (SA: $\overline{\Gamma}_1 = 150.9$~Mbps; MA: $\overline{\Gamma}_1 = 149.7$~Mbps), demonstrating the advantages of contextual methods in dynamic scenarios. Both SA and MA variants reliably adapt to interval-level dynamics by consistently converging to the optimal channel in each interval, as observed for the representative trial: channel~\#1:$\{1\}$ is selected in $99.7\%$ and $100\%$ of decisions during the first interval, \#2:$\{2\}$ in $99.7\%$ and $100\%$ during the second, \#3:$\{3\}$ in $97.4\%$ and $99.9\%$ during the third, and \#1:$\{1\}$ again in $99.9\%$ and $100\%$ during the fourth, for SA and MA, respectively. Again, as shown in Fig.~\ref{fig:b_evo}, MA~LinUCB converges faster than its SA counterpart, a recurring pattern across trials. Notably, LinUCB converges especially quickly in the fourth interval, where a previously optimal channel (e.g., \#1:$\{1\}$ in the representative trial) reappears as lightly traffic-loaded, since contextual information enables immediate re-identification rather than renewed exploration, confirming its efficiency in recurrent conditions. This behavior is consistently observed for LinUCB across trials, as indicated by its high frequency of optimal selections in the fourth interval, whereas UCB lacks such consistency. Indeed, although SA~UCB also converges rapidly in the representative trial, in other trials it selects the fourth-interval optimum in only about $70\%$ of decisions, while MA~UCB often fails to identify it. In contrast, SA~LinUCB and MA~LinUCB consistently select the optimal channel in over $96\%$ and $99\%$ of decisions, respectively.

\noindent\textit{\textbf{Discussion:}}
Scenario~SP2 highlights the benefits of learning under dynamic conditions, as agents consistently outperform static allocations by exploiting temporarily underloaded channels. UCB achieves strong performance, but MA~UCB remains susceptible to the previously identified coordination issues. LinUCB again provides the highest performance, reliably adapting to traffic dynamics and recurrent patterns thanks to using relevant contextual information that prevents relearning deployment conditions. Across algorithms, MA settings achieve comparable performance to SA, while converging faster. 

\section{Results: Multi-Player Game}\label{sec:results_2}

This section presents the simulation results for the MP game under SCB. For benchmarking, the baseline assumes all APs are legacy, non-learning devices statically configured to operate over an 80~MHz channel, representing a typical OBSS scenario.

\subsection{Learning under full-buffer conditions (Scenario~MP1)}

\begin{figure}[t!]
\centering
    \begin{subfigure}[t]{\linewidth}
        \centering
        \includegraphics[width=\linewidth]{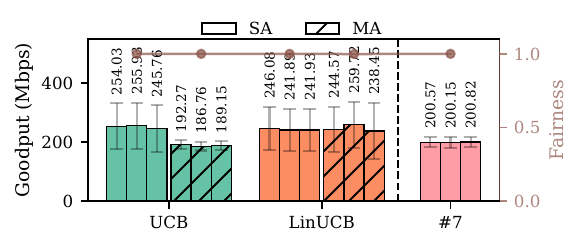}
        \caption{Scenario~MP1 (sub-bars correspond to BSS$_1$--BSS$_3$).}
        \label{fig:res_conf1}
    \end{subfigure}%
    \hfill
    \begin{subfigure}[t]{\linewidth}
        \centering
        \includegraphics[width=\linewidth]{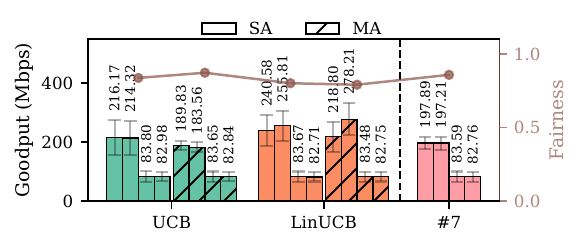}
        \caption{Scenario~MP2 (sub-bars correspond to BSS$_1$--BSS$_4$).}
        \label{fig:res_conf2}
    \end{subfigure}
    \begin{subfigure}[t]{\linewidth}
        \centering
        \includegraphics[width=\linewidth]{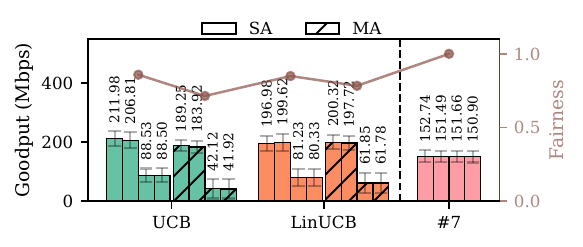}
        \caption{Scenario~MP3 (sub-bars correspond to BSS$_1$--BSS$_4$).}
        \label{fig:res_conf3}
    \end{subfigure}
    \caption{Goodput results across all trials under SCB, comparing each algorithm--architecture pair and the static non-learning baseline. Brown markers indicate goodput fairness computed across all BSSs.}
    \label{fig:res_conf_combined}
\end{figure}

Fig.~\ref{fig:res_conf1} reports the goodput experienced by each BSS (i.e., BSS$_1$, BSS$_2$, and BSS$_3$) across all trials under the different learning approaches and IEEE~802.11 baseline, including performance fairness across BSSs.

\noindent\textit{\textbf{Learning algorithms:}}  
Learning algorithms exhibit distinct allocation patterns that can be classified as:  
(i) two BSSs on the same 40~MHz channel, with the remaining BSS on the other 40~MHz channel;  
(ii) each BSS on a distinct 20~MHz channel;  
(iii) one BSS on a 40~MHz channel, and the other two on separate, contention-free 20~MHz channels. 

In particular, UCB shows distinct allocation preferences between the SA and MA variants. SA~UCB predominantly converges to~(i), achieving high average goodput across trials but reduced per-trial fairness, as the BSS operating alone on a 40~MHz channel systematically attains superior performance (e.g., in a representative trial: $\overline{\Gamma}_1 = 227.7$~Mbps, $\overline{\Gamma}_2 = 213.2$~Mbps, and $\overline{\Gamma}_3 = 303.7$~Mbps, leading to $\mathcal{J}=0.975$). 
In contrast, MA~UCB consistently selects~(ii), a conservative strategy that avoids overlap but underutilizes bandwidth, yielding lower average goodputs while sustaining fairness (e.g., in a representative trial: $\overline{\Gamma}_1 = 189.0$~Mbps, $\overline{\Gamma}_2 = 184.2$~Mbps, $\overline{\Gamma}_3 = 189.0$~Mbps, leading to $\mathcal{J}\approx 1$).

On the other hand, LinUCB also exhibits distinct preferences between architectures. SA~LinUCB consistently adopts~(i), resulting in high goodputs across BSSs but reduced fairness (e.g., in a representative trial: $\overline{\Gamma}_1 = 310.7$~Mbps, $\overline{\Gamma}_2 = 207.4$~Mbps, $\overline{\Gamma}_3 = 244.4$~Mbps, leading to $\mathcal{J}=0.973$).
In contrast, MA~LinUCB converges to~(iii), achieving high but uneven goodput among BSSs (e.g., in a representative trial: $\overline{\Gamma}_1 = 357.2$~Mbps, $\overline{\Gamma}_2 = 207.5$~Mbps, $\overline{\Gamma}_3 = 192.3$~Mbps, leading to $\mathcal{J}=0.920$).  

\noindent\textit{\textbf{Discussion:}} 
All learning algorithms consistently avoid 80~MHz allocations, thereby mitigating excessive contention-induced queuing delays. They demonstrate effective distribution of channels across BSSs, though MA~UCB sacrifices bandwidth efficiency, likely due to the previously identified joint-action coupling problem. Nevertheless, as illustrated in Fig.~\ref{fig:res_conf1}, all other learning approaches---despite exhibiting distinct allocation preferences---consistently outperform the static baseline, which achieves approximately $200$~Mbps per BSS across trials. This diversity in learned behaviors arises from the non-convex action-reward landscape, which contains multiple local optima, causing independent learners to stabilize at different equilibrium configurations based on their exploration and initialization.
Among the learning approaches, LinUCB achieves the highest performance. Both, SA and MA LinUCB achieve comparable steady-state goodputs, yielding average gains across all BSSs of $+21.3\%$ and $+23.5\%$ over the baseline, respectively. However, when accounting for the initial learning phase, SA~LinUCB's gain drops to $+17.2\%$ due to slower convergence, whereas MA~LinUCB maintains a $+21.3\%$ improvement, highlighting the benefits of decomposing the learning process across multiple agents. 

\subsection{Learning under heterogeneous load distributions (Scenario~MP2)}

\begin{figure}[t!]
    \centering
    \begin{subfigure}[t]{\linewidth}
        \centering
        \includegraphics[width=\linewidth]{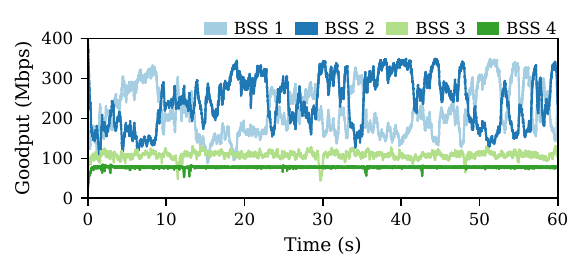}
        \caption{SA LinUCB.}
        \label{fig:2_evo_linucb}
    \end{subfigure}%
    \caption{Temporal evolution of goodput for each BSS in a representative trial from Scenario~MP2 under SCB. Target offered loads for BSS$_3$ and BSS$_4$: $108.5$ and $79.3$~Mbps, respectively.}
    \label{fig:2_evo}
\end{figure}

Fig.~\ref{fig:res_conf2} reports the goodput experienced by each BSS (i.e., BSS$_1$ to BSS$_4$) across all trials under the different learning approaches and IEEE~802.11 baseline, including performance fairness across BSSs. 
Fig.~\ref{fig:2_evo_linucb} further illustrates the temporal evolution of BSS$_1$ goodput for a learning approach (SA~LinUCB) in a representative trial.

\noindent\textit{\textbf{Learning algorithms:}} 
Learning algorithms again exhibit distinct allocation preferences under heterogeneous traffic:  
(i) both high-traffic BSSs on the same 40~MHz channel, with low-traffic BSSs on separate contention-free 20~MHz channels;  
(ii) each BSS on a distinct 20~MHz channel;  
(iii) a pair of high-traffic and low-traffic BSSs on distinct 40~MHz channels;  
(iv) both high-traffic BSSs on distinct 40~MHz channels, each partially overlapped by a 20~MHz low-traffic BSS;  
(v) hybrid allocations, where one high-traffic BSS shares a 40~MHz channel with a low-traffic BSS, while the others occupy non-overlapping, contention-free 20~MHz channels. 

In particular, UCB predominantly converges to~(i) in its SA variant. As illustrated in Fig.~\ref{fig:res_conf2}, this yields high fairness among high-traffic BSSs, balancing their performance (e.g., in the representative trial: $\overline{\Gamma}_1 = 202.7$~Mbps and $\overline{\Gamma}_2 = 207.4$~Mbps, leading to $\mathcal{J}=0.999$). Conversely, MA~UCB, as in Scenario~ML1, consistently converges to~(ii), a conservative allocation that penalizes high-traffic BSSs due to limited bandwidth. The previous identified coupling problem emerges again: BSSs allocated to higher-index channels in the action space suffer greater degradation due to association with larger CWs (e.g., in the representative trial: BSS$_1$ on \#3:$\{3\}$ achieves $180.9$~Mbps, while BSS$_2$ on \#1:$\{1\}$ reaches $198.2$~Mbps). 

On the other hand, LinUCB alternates across trials between policies~(iii) and (iv) in its SA variant, thereby consistently placing the two high-traffic BSSs on distinct 40~MHz channels. 
As illustrated in Fig.~\ref{fig:res_conf2}, this strategy yields the highest average performance for high-traffic BSSs across trials, while ensuring low-traffic BSSs modest demands are met. As shown in Fig.~\ref{fig:2_evo_linucb} for the representative trial, low-traffic BSSs achieve stable goodput near their targets ($\overline{\Gamma}_3 = 108.1$~Mbps, $\overline{\Gamma}_4 = 78.2$~Mbps), whereas high-traffic BSSs exhibit oscillatory behavior: when one improves, the other degrades proportionally. This ``ping-pong" effect reflects the greedy and reactive nature of learning APs. On the other hand, MA~LinUCB alternates across trials among~(i), (iii), and (v), achieving strong overall performance, as illustrated in Fig.~\ref{fig:res_conf2}. Nevertheless, under allocation~(v), it exhibits reduced fairness across high-traffic BSSs (e.g., in the  representative trial under allocation~(v): BSS$_1$ achieves $194.2$~Mbps using 20~MHz, while BSS$_2$ achieves $289.1$~Mbps using 40~MHz, yielding $\mathcal{J}=0.963$).

\noindent\textit{\textbf{Discussion:}}
As in Scenario~MP1, all learning algorithms avoid 80~MHz allocations, which, despite preserving fairness, suffer severe contention-induced degradation. Thereby, as shown in Fig.~\ref{fig:res_conf2}, learning agents outperform the baseline for high-traffic BSSs---except MA~UCB, which remains limited by its coupling problem. Low-traffic BSSs, in contrast, reliably satisfy their modest demands across all algorithms and the baseline, consistently achieving over $95\%$ of their targets.  
LinUCB achieves again the most effective adaptation: SA~LinUCB delivers a $+25.6\%$ gain on high-traffic BSSs over the baseline ($+21.8\%$ including the learning phase), while MA~LinUCB achieves $+25.7\%$ ($+24.5\%$ including the learning phase), demonstrating the advantages of contextual strategies. Finally, the results demonstrate that selfish multi-player settings may exhibit oscillatory behavior, as decentralized interactions can induce instability and policy-chasing dynamics.

\subsection{Learning vs. Legacy coexistence (Scenario~MP3)}

\begin{figure}[t!]
    \centering
    \begin{subfigure}[t]{\linewidth}
        \centering
        \includegraphics[width=\linewidth]{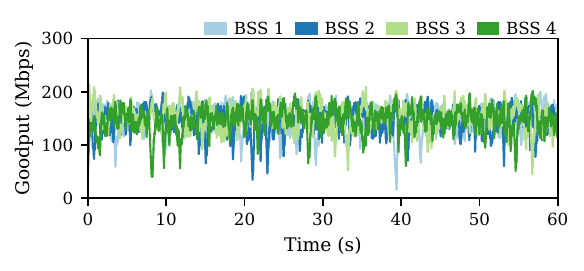}
        \caption{Baseline (\#7:$\{1,2,3,4\})$.}
        \label{fig:3_evo_baseline}
    \end{subfigure}
    \begin{subfigure}[t]{\linewidth}
        \centering
        \includegraphics[width=\linewidth]{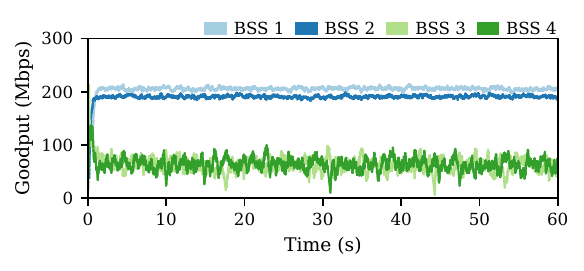}
        \caption{SA LinUCB.}
        \label{fig:3_evo_linucb}
    \end{subfigure}%
    \caption{Temporal evolution of goodput for each BSS in a
representative trial from Scenario~MP3 under SCB. Target
offered loads for BSS$_1$--BSS$_4$: ~$208.0$, $192.9$, $248.8$, and $205.0$~Mbps, respectively.}
    \label{fig:3_evo}
\end{figure}

Fig.~\ref{fig:res_conf3} reports the goodput experienced by each BSS (i.e., BSS$_1$ to BSS$_4$) across all trials under the different learning approaches and IEEE~802.11 baseline, including performance fairness across BSSs. Figures.~\ref{fig:3_evo_baseline} and ~\ref{fig:3_evo_linucb} further illustrate the temporal evolution of goodput per BSS in a representative trial for the baseline and a learning approach (SA~LinUCB), respectively.

\noindent\textit{\textbf{Learning algorithms:}} 
Learning algorithms again exhibit distinct allocation preferences in the presence of legacy 80~MHz BSSs:  
(i) each learning BSSs on a separate 40~MHz channels;  
(ii) each learning BSS on a distinct 20~MHz channel; 
(iii) both learning BSSs on the 80~MHz channel;  
(iv) one learning BSS on a 40~MHz channel and the other on a non-overlapping 20~MHz channel. 
In allocation~(i), under SCB, learning BSSs benefit from increased channel access opportunities, since legacy BSSs defer unless the entire band is idle. This boosts learning BSSs’ goodput but severely degrades legacy performance.  
In allocation~(ii), compared to (i), both learning and legacy BSSs are penalized, provided that learning BSSs suffer reduced transmission efficiency due to narrower channels and legacy BSSs experience increased contention from smaller-bandwidth contenders.  
In allocation~(iii), corresponding to the baseline allocations, learning and legacy BSSs achieve comparable and balanced performance, improving fairness.  
Finally, allocation~(iv) yields asymmetric allocations among learning BSSs, reducing intra-learning fairness. 

Regarding the performance of learning solutions, UCB alternates across trials between~(i) and~(iii) in its SA variant (e.g., in the representative trial under allocation~(i): $\overline{\Gamma}_1 = 199.9$~Mbps and $\overline{\Gamma}_2 = 190.1$~Mbps, while legacy BSSs drop to $\overline{\Gamma}_3 =81.6$ and $\overline{\Gamma}_4 =76.5$~Mbps, yielding $\mathcal{J}=0.848$). MA~UCB again predominantly selects allocation~(ii), significantly reducing overall performance, as illustrated in Fig.~\ref{fig:res_conf3} (e.g., in the representative trial: learning BSSs achieve $\overline{\Gamma}_1 =188.0$ and $\overline{\Gamma}_2 =190.7$~Mbps, while legacy BSSs collapse to $\overline{\Gamma}_3 =26.1$ and $\overline{\Gamma}_4 =24.7$~Mbps, yielding $\mathcal{J}=0.632$).

On the other hand, LinUCB alternates across trials between~(i) and~(iii) in its SA variant. As shown in Fig.~\ref{fig:3_evo_linucb}, in the representative trial under allocation~(i), learning BSSs achieve their offered loads ($\overline{\Gamma}_1 = 206.1$~Mbps, $\overline{\Gamma}_2 = 191.1$~Mbps), while legacy BSSs are severely degraded ($\overline{\Gamma}_3 =64.2$ and $\overline{\Gamma}_4 =65.3$~Mbps). In contrast, as shown in Fig.~\ref{fig:3_evo_baseline}, under baseline operation, goodputs remain balanced but unstable ($\overline{\Gamma}_1 = 152.9$ and $\overline{\Gamma}_2 = 149.6$ Mbps for learning BSSs, and $\overline{\Gamma}_3 = 155.7$ and $\overline{\Gamma}_4 = 150.8$~Mbps for legacy ones, yielding $\mathcal{J}\approx1$), and none achieves its target load. On the other hand, MA~LinUCB alternates across trials between ~(i), (ii), and~(iv) (e.g., in the representative trial under allocation~(ii): learning BSSs sustain $\overline{\Gamma}_1 = 186.8$ and $\overline{\Gamma}_2 = 186.4$~Mbps, while legacy BSSs drop to $\overline{\Gamma}_3 = 31.3$ and $\overline{\Gamma}_3 = 31.4$~Mbps, yielding $\mathcal{J}=0.663$). Again both SA and MA variants achieve comparable performance for learning BSSs, as illustrated in Fig.~\ref{fig:res_conf3}. Nevertheless, MA preferences have greater impact on legacy BSSs.

\noindent\textit{\textbf{Discussion:}}
All learning algorithms effectively avoid placing the two learning BSSs on overlapping 20/40~MHz channels. As shown in Fig.~\ref{fig:res_conf3}, all methods significantly increase the goodput of learning BSSs relative to the baseline. However, these gains consistently come at the expense of legacy BSSs, which suffer from persistent contention with narrower-bandwidth contenders and overlap on secondary channels. 
Importantly, under baseline operation, the goodput of all BSSs remains below $90\%$ of their offered load in every trial. Learning algorithms, however, enable the learning BSSs to reach this threshold in a substantial fraction of trials (e.g., SA~UCB: in $90\%$ for both BSS$_1$ and BSS$_2$; SA~LinUCB: in $50\%$ and $80\%$, respectively). 
Again, MA~UCB performs worst overall, as its conservative allocation~(ii) allocations reduce efficiency for both learning and legacy BSSs. In contrast, LinUCB achieves the highest gains for the learning BSSs, at the expense of the legacy BSSs. For instance, MA~LinUCB increases the goodput of the learning BSSs by $+30.4\%$ relative to the baseline, but results in a $59.1\%$ loss for the legacy BSSs.

\section{Is dynamic channel bonding changing things?}\label{sec:results_3}

This section evaluates the impact of enabling DCB in both SP and MP settings, using Scenario~SP2 and Scenario~MP1 as representative cases. DCB introduces greater flexibility as it allows APs to adapt their transmission bandwidth to instantaneous channel availability, increasing spectrum efficiency and reducing contention.

\subsection{Dynamic OBSS traffic (Scenario~SP2)}

\begin{figure}[t!]
\centering
    \begin{subfigure}[t]{\linewidth}
        \centering
        \includegraphics[width=\linewidth]{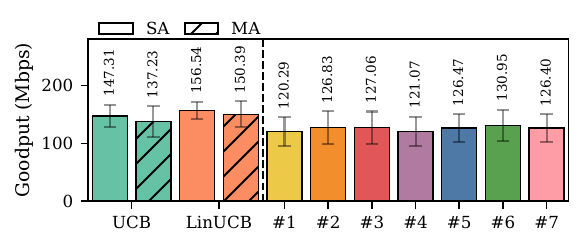}
        \caption{Aggregate across trials.}
        \label{fig:res_b_goodput_dcb}
    \end{subfigure}%
    \hfill
    \begin{subfigure}[t]{\linewidth}
        \centering
        \includegraphics[width=\linewidth]{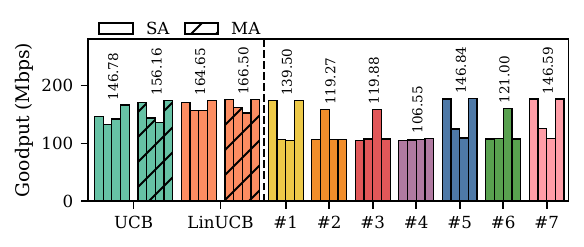}
        \caption{Representative trial (sub-bars correspond to 15-second intervals).}
        \label{fig:res_b_goodput_intervals_dcb}
    \end{subfigure}
    \caption{Scenario~SP2 goodput results for BSS$_1$ under DCB, comparing each algorithm--architecture pair and static non-learning baseline.}
    \label{fig:res_b_combined2}
\end{figure}

Fig.~\ref{fig:res_b_goodput_dcb} reports the goodput achieved by the different learning approaches and IEEE~802.11 baselines across all trials. In contrast, Fig.~\ref{fig:res_b_goodput_intervals_dcb} presents the goodput observed in each 15-second interval of the simulation for a representative trial. Recall that, in the considered baseline configurations, the primary channel is assigned to the lowest-indexed basic 20~MHz channel within each configured operational channel. Other primary channels may yield distinct performances.

\noindent \textit{\textbf{IEEE 802.11 legacy operation:}} 
As illustrated in Fig.~\ref{fig:res_b_goodput_dcb} (right), under DCB, 40~MHz and 80~MHz allocations achieve performance comparable to 20~MHz configurations. This occurs because these wider allocations enjoy similar channel access opportunities, transmitting whenever their primary channel is idle (i.e., channel~$1$ for \#5:$\{1p,2\}$ and \#7:$\{1p,2,3,4\}$, channel~$3$ for \#6:$\{3p,4\}$). They even achieve slightly higher goodput than the corresponding single channel allocations, as they occasionally transmit at increased bandwidths, although opportunities remain limited due to overlapping BSS loads.
As illustrated in Fig.~\ref{fig:res_b_goodput_intervals_dcb} (right) for the same representative trial as in the SCB case, 40~MHz and 80~MHz allocations become best-performing when their primary channel is underloaded (e.g., \#6:$\{3p,4\}$ in the third interval: $\overline{\Gamma}_1 = 160.7$~Mbps), achieving results comparable to the best 20~MHz allocations. In other intervals, their performance remains similar to any 20~MHz configuration not mapped to the underloaded channel. These results confirm that, under DCB, the primary channel choice is critical, as it directly determines access opportunities and overall performance.

\noindent\textit{\textbf{Learning algorithms:}} 
As shown in Fig.~\ref{fig:res_b_goodput_dcb} (left), learning algorithms achieve high performance under DCB, comparable to that observed under SCB. Specifically, under DCB, the algorithms consistently prioritize configurations using the underloaded channel as the primary. \\
In particular, UCB in its SA variant distributes selections across configurations using the underloaded channel as primary, reflecting the presence of multiple local optima. For instance, in the representative trial, SA~UCB selects during the first interval \#1:$\{1p\}$ in $30.6\%$ of decisions, \#5:$\{1p,2\}$ in $29.5\%$, and \#7:$\{1p,2,3,4\}$ $33.8\%$, and during the third interval, it selects \#3:$\{3p\}$ in $30.4\%$, \#6:$\{3p,4\}$ in $32.5\%$, and \#7$_{_3}$:$\{1,2,3p,4\}$ in $31.7\%$. 
In contrast, MA~UCB consistently converges to the optimal primary-only 20~MHz allocations across intervals. Interestingly, as illustrated in Fig.~\ref{fig:res_b_goodput_intervals_dcb} (left), its performance under DCB significantly surpasses SCB in the representative trial, as in the third interval it correctly identifies channel~\#3:$\{3p\}$ (selected in $86.7\%$ of decisions) as optimal, whereas under SCB it remained on \#1:$\{1p\}$. This highlights the stronger guiding role of primary channel selection under DCB.

On the other hand, as illustrated in Fig.~\ref{fig:res_b_goodput_dcb} (left), LinUCB achieves the highest overall performance across trials. Both SA and MA variants distribute selections across configurations using the interval-wise optimal primary. For instance, in the representative trial, SA~LinUCB selects during the first interval \#1:$\{1p\}$ in $22.1\%$ of decisions, \#5:$\{1p,2\}$ in $62.2\%$, and \#7:$\{1p,2,3,4\}$ in $16.2\%$, whereas during the third interval, it selects \#3:$\{3p\}$ in $35.3\%$ of decisions and \#5:$\{3p,4\}$ in $63.4\%$. It is noteworthy that MA~LinUCB tends to avoid the 80~MHz allocation more often than other allocations, likely due to the larger (masked) action space of the primary channel agent (e.g., $|\mathcal{A}^{\text{p}}| = 4$ versus $|\mathcal{A}^{\text{p}}| = 1$ for an 80-Mhz and a 20-MHz operational channel, respectively).

\noindent\textit{\textbf{Discussion:}}
Learning algorithms demonstrate effective learning under DCB, as they consistently converge to configurations whose primary channel is underloaded, exploiting DCB's flexibility and generally surpassing baseline performance. Indeed, under DCB, primary channel selection becomes critical, as it enables wider allocations to perform competitively.  

\subsection{Learning under full-buffer conditions (Scenario~MP1)}
\begin{figure}[t!]
    \centering
    \begin{subfigure}[t]{\linewidth}
        \centering
        \includegraphics[width=\linewidth]{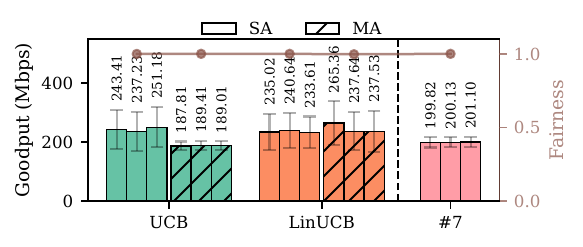}
        \caption{Scenario~MP1 (sub-bars correspond to BSS$_1$--BSS$_3$).}
        \label{fig:}
    \end{subfigure}%
    \caption{Goodput results across all trials under DCB, comparing each algorithm--architecture pair and the static non-learning baseline. Brown markers indicate goodput fairness computed across all BSSs.}
    \label{fig:res_conf1_dcb}
\end{figure}

Fig.~\ref{fig:res_conf1_dcb} reports the goodput experienced by each BSS across all trials under DCF for the different learning approaches and IEEE 802.11 baseline, including performance fairness across BSSs.

\noindent\textit{\textbf{Learning algorithms:}}  
Under DCB, BSSs can opportunistically exploit available secondary channels beyond their primary one, increasing transmission opportunities and effective bandwidth. Such flexibility enables multiple favorable allocation configurations and, consequently, learning algorithms exhibit more diverse allocations than under SCB:  
(i) all BSSs on the 80~MHz channel;  
(ii) each BSS on a distinct 20~MHz channel;  
(iii) two BSSs on the 80~MHz channel and the remaining BSS on a 40~MHz channel;  
(iv) two BSSs on separate 40~MHz channels and the remaining BSS on the 80~MHz channel;  
(v) two BSSs on separate 40~MHz channels and the remaining BSS on a 20~MHz channel;  
(vi) one BSS on 80~MHz, one on 40~MHz, and the remaining BSS on a 20~MHz channel overlapping with the 40~MHz one. Nevertheless, learning algorithms consistently assign a distinct primary channel to each BSS, enabling medium access through the primary whenever it is idle, even if some secondary channels are busy. 

In particular, UCB again achieves higher average goodput in the SA variant than in the MA variant across trials (SA: $\overline{\Gamma}_1 = 243.4$, $\overline{\Gamma}_2 = 237.2$, and $\overline{\Gamma}_3 = 251.2$~Mbps; MA: $\overline{\Gamma}_1 = 187.8$, $\overline{\Gamma}_2 = 189.4$, and $\overline{\Gamma}_3 = 189.0$~Mbps). SA~UCB primarily converges to~(iii), but occasionally selects~(iv). Notably, under allocation~(iii), SA~UCB consistently assigns a primary channel for the 80~MHz BSSs that is not included within the 40~MHz channel of the other BSS, thereby reducing direct contention. Conversely, MA~UCB again consistently selects~(ii) due to its coupling problem, resulting in persistently conservative 20~MHz allocations that forgo the benefits of DCB.

On the other hand, LinUCB achieves high performance across trials in both architectures. SA~LinUCB primarily converges to~(iii), but occasionally selects~(i), yielding average goodputs comparable to SA~UCB ($\overline{\Gamma}_1 = 235.0$, $\overline{\Gamma}_2 = 240.6$, and $\overline{\Gamma}_3 = 233.6$~Mbps). Conversely, MA~LinUCB alternates between~(v) and (vi), achieving high goodputs across trials ($\overline{\Gamma}_1 = 265.4$, $\overline{\Gamma}_2 = 237.2$, and $\overline{\Gamma}_3 = 237.5$~Mbps). Having BSSs operating with distinct yet partially overlapping effective bandwidths may enable BSSs to better exploit the flexibility of DCB and access additional transmission opportunities, albeit at the expense of fairness.

\noindent\textit{\textbf{Discussion:}}  
The baseline configuration (in which all BSSs operate on the 80~MHz channel and share the same primary) limits DCB's flexibility, as medium access depends solely on 80~MHz availability without the fallback to smaller channels. Consequently, it achieves comparable performance under DCB and SCB (around $200$~Mbps per BSS). In contrast, learning algorithms (except MA~UCB) consistently outperform the baseline by enabling BSSs to opportunistically exploit narrower bandwidths, but at the expense of fairness. 
Indeed, despite exhibiting distinct allocation preferences, all learning approaches demonstrate implicit coordination in multi-player settings by assigning distinct primary channels to BSSs and harnessing the benefits of DCB.

\section{Discussion and Takeaways}\label{sec:discussion}

Our results confirm that (C)MAB-based learning provides a practical and effective alternative to legacy IEEE~802.11 static channel access, offering adaptability to environmental conditions and mitigating the adverse effects of misconfiguration under both SCB and DCB. Remarkably, even in multi-player scenarios, learning-enabled APs achieve implicit coordination without explicit communication, driven by their mutual interests as discussed in~\cite{barrachina2021multi}; however, their inherent greediness can reduce fairness, degrade coexisting legacy BSSs’ performance, and induce oscillatory policy-chasing behaviors.

Among the evaluated strategies, both SA and MA~LinUCB consistently achieved the highest performance, significantly surpassing UCB, highlighting the benefits of contextual learning, as also observed in~\cite{martinez2023contextual} for network selection.
In addition to superior steady-state performance, LinUCB also demonstrated faster convergence under recurrent conditions, as contextual information enables immediate re-identification of previously encountered situations rather than renewed exploration.
SA~UCB delivered reasonable performance, but MA~UCB consistently suffered from coupling problems. In contrast, MA~LinUCB avoided this limitation because each agent conditions its action selection on independent contextual information. Hierarchical learning---adopted in~\cite{wojnar2024ieee, wojnar2025coordinated}---offers a potential remedy to MA~UCB’s limitations by decoupling decisions across dimensions, but its scalability is limited 
 in high-dimensional spaces.  

Learning architecture strongly influenced convergence and robustness.
MA strategies generally converged faster than SA approaches, likely because each agent in the MA setting explores a smaller, decoupled action space (performing component-wise optimization), while the agent in the SA setting faces a larger combinatorial joint action space. Moreover, under LinUCB, both architectures achieved comparable steady-state performance, highlighting the practical potential of MA designs. However, as shown for UCB, MA settings require proper coordination among agents to avoid inefficiencies such as credit assignment problems.

Thus, the main takeaways of this paper can be summarized as follows:  
\begin{itemize}  
    \item (C)MAB-based learning offers a robust alternative to static IEEE~802.11 configurations, which often operate suboptimally and fail to adapt to dynamic network conditions.
    \item Contextual learning, particularly LinUCB, consistently outperforms non-contextual methods by leveraging environmental features; non-contextual approaches remain simpler but may face inherent structural limitations. 
    \item Beyond enabling modular and flexible systems, MA approaches converge significantly faster than SA while achieving comparable performance under LinUCB. 
    \item In multi-player settings, learning enables implicit coordination among APs, yet their greediness significantly affects fairness and can trigger oscillatory policy-chasing dynamics, highlighting the need for inter-AP coordination mechanisms.  
\end{itemize}  

\section{Conclusions} \label{sec:conclusions}

In this work, we evaluated SA and MA (C)MAB-based learning architectures for MAC-layer optimization (jointly selecting channel, primary channel, and contention window size) across both single- and multi-player games. Our assessment included contextual (LinUCB) and non-contextual (UCB) state-of-the-art MAB methods.  

Future work may extend the action space to additional MAC and PHY parameters, including enabling/disabling A-MPDU frame aggregation and RTS/CTS protection, or transmission power and MCS selection, potentially introducing instability in uncoordinated MA architectures due to oscillations driven by interdependent decisions. Another promising direction is investigating how the dimension of the action space affects learning performance in SA and MA settings, potentially revealing advantages of distributing complex decision-making across multiple agents. Enhancing context observability through signal quality or interference measurements could improve policy adaptation, while alternative reward formulations could promote fairness, maximize throughput, or minimize airtime. Also, investigating stateful RL approaches may capture the long-term consequences of agent decisions more effectively than bandit-based methods. Finally, exploring scenarios with mobility and heterogeneous multi-learning deployments could provide further insights into learning dynamics.


\section{Acknowledgments}

This paper is supported by the CHIST-ERA Wireless AI 2022 call MLDR project (ANR-23-CHR4-0005), partially funded by AEI and NCN under projects PCI2023-145958-2 and 2023/05/Y/ST7/00004, respectively, by Wi-XR PID2021-123995NB-I00 and TRUE-Wi-Fi PID2024-155470NB-I00 (MCIU/AEI/FEDER,UE), by MCIN/AEI under the Maria de Maeztu Units of Excellence Programme (CEX2021-001195-M), and AGAUR ICREA Academia 00077.


\bibliographystyle{unsrt} 
\bibliography{references}

\end{document}